\documentclass[a4paper,11pt]{article}
\pdfoutput=1 % if your are submitting a pdflatex (i.e. if you have
\usepackage{jheppub} % for details on the use of the package, please
\usepackage[T1]{fontenc} % if needed
\usepackage{blkarray}
\usepackage{relsize}
\usepackage{slashed}
\usepackage{filecontents}
\usepackage{mathtools}

\def \barlambda {\bar{\lambda}}
\def \barmu {\bar{\mu}}

\allowdisplaybreaks

\title{\boldmath A Pure Spinor Twistor Description of the $D=10$ Superparticle}

\author[a,b]{Diego Garc\'ia Sep\'ulveda}
\author[c]{and Max Guillen}

\affiliation[a]{Perimeter Institute for Theoretical Physics, Waterloo, ON N2L 2Y5, Canada}
\affiliation[b]{
ICTP South American Institute for Fundamental Research \\
Instituto de F\'isica Te\'orica, UNESP-Universidade Estadual Paulista \\
R. Dr. Bento T. Ferraz 271, Bl. II, S\~ao Paulo 01140-070, SP, Brazil}
\affiliation[c]{Department of Physics and Astronomy,\\
Uppsala University, 75108 Uppsala, Sweden}

\emailAdd{diego.garcia@unesp.br}
\emailAdd{max.guillen@physics.uu.se}

\abstract{We present a novel twistor formulation of the ten-dimensional massless superparticle. This formulation is based on the introduction of pure spinor variables through a field redefinition of another model for the superparticle, and in the new description we find that the super-Pauli-Lubanski three-form naturally arises as a constraint. Quantization is studied in detail for both models and they are shown to correctly describe the $D=10$ super-Yang-Mills states.}

\begin{document} 
\maketitle

\flushbottom

\section{Introduction} \label{introduction}

After the advent of the twistor string construction \cite{Witten:2003nn, Berkovits:2004hg}, twistor techniques and formulas in four dimensional field theory were rapidly developed \cite{Britto:2004ap, Britto:2005fq, Roiban:2004vt} and have proven to be a powerful toolbox for computing scattering amplitudes and thus understanding quantum field theory in a more efficient way compared to traditional approaches. In this regard, special advances have been uncovered for $\mathcal{N}=4$, $D=4$ super-Yang-Mills theory, and since such theory can be obtained from $D=10$ super-Yang-Mills via dimensional reduction, it is tempting to ask if similar twistor constructions exist in higher dimensions. In this respect, pure spinors have been argued to be the natural extension of twistors to higher dimensions, whether by trying to generalize Penrose's original construction in four dimensions \cite{doi:10.1063/1.1705200, Hughston1, Hughston2, Hughston3, Berkovits:2004bw, Boels:2009bv}, or by considering the relation between integrability along pure spinor/light-like lines and super-Yang-Mills \cite{Witten:1985nt, Howe:1991mf}. Pure spinors have also been shown to be useful for covariant quantization of superstrings in a manifestly supersymmetric way \cite{Berkovits:2000fe}, and has allowed much progress in the computation and understanding of string amplitudes \cite{Mafra:2011nv, Mafra:2011nw, Mafra:2018nla, Mafra:2018pll, Mafra:2018qqe, Gomez:2013sla}. Thus, it becomes promising to consider pure spinors to try to generalize the results of the original four-dimensional twistor string program to ten dimensions, where superstring theory naturally lives.

In resonance with these ideas, Berkovits attempted in \cite{Berkovits:2009by} a twistor-like construction in ten dimensions from which the three-point super-Yang-Mills ``stripped amplitude''\footnote{By which we mean a scattering amplitude without considering any delta functions.} was recovered and some relations to the standard pure spinor formalism were pointed out. The construction relies on the use of the supertwistor variables
\begin{equation} \label{nathanvariablesin}
    \mathcal{Z}^{I} = (\lambda^{\alpha},\, \mu_{\alpha}, \,\Gamma^{m}), \ \ \ \ \ \bar{\mathcal{Z}}_{I} = (\bar{\mu}_{\alpha}, \, -\bar{\lambda}^{\alpha}, \, \bar{\Gamma}^{m}),
\end{equation}
where $\lambda^{\alpha}$ is a pure spinor, $\bar{\lambda}^{\alpha}$ is a 16-component spinor, and $\Gamma^{m}$ is a fermionic vector. The variables defining $\bar{\mathcal{Z}}_{I}$ correspond to the canonical conjugates to the variables defining $\mathcal{Z}^{I}$, and by definition they are required to solve the constraints
\begin{equation}
    \lambda \mu = 0, \ \ \ (\lambda \gamma^{mn} \mu) + 4 \Gamma^{m} \Gamma^{n} = 0, \ \ \ \Gamma^{m}(\lambda \gamma_{m})_{\alpha} = 0.
\end{equation}

Despite the former achievements, the twistor-like construction as strictly developed in \cite{Berkovits:2009by} lacked an action principle, and thus some of the ingredients were just proposed using insights from ordinary pure spinor strings. Finding an origin for these elements from first-principles was then left as an open problem.

In this work we will present a model for the ten-dimensional massless superparticle \cite{Brink:1981nb} -which we call the pure spinor twistor superparticle- that readily makes use of the pure spinor twistor variables \eqref{nathanvariablesin}, and naturally incorporates the elements introduced in \cite{Berkovits:2009by}. For this purpose, we start from yet another model of the superparticle developed by Berkovits in \cite{Berkovits:1990yc} and establish a consistent map between the two models. The Berkovits' superparticle construction is based on the introduction of a spinor $\Lambda^{\alpha}$ which solves the massless condition $P^{2}=0$ as $P^{m} = \Lambda \gamma^{m} \Lambda$ in virtue of the special ten-dimensional identity $(\gamma^{m})_{\alpha (\beta} (\gamma_{m})_{\gamma \delta)} = 0$. The spinor $\Lambda^{\alpha}$ is constrained in order to reproduce the correct number of degrees of freedom for $P^{m}$. Before presenting the relation between the two models, we study the canonical quantization and BRST quantization of Berkovits' model. The classical bosonic piece of the BRST operator was effectively worked out in \cite{Carabine:2018kdg}, and here we will present the full quantum supersymmetric BRST operator and quantization.

As a result of the relation between the two superparticle descriptions, we will see that the constraints that the pure spinor twistor superparticle must satisfy are
\begin{align} 
    & J \coloneqq  \barmu \lambda - \barlambda \mu + \bar{\Gamma}^{m}\Gamma_{m} = 0, \label{purespinortwistorconstraints1} \\[1.0ex] &B \coloneqq \frac{1}{2} \Big[ (\barlambda \gamma^{m} \barlambda) \Gamma_{m} - (\lambda \gamma^{m} \barlambda) \bar{\Gamma}_{m} \Big] = 0,  \label{purespinortwistorconstraints2}\\[1.0ex]
    & 
    \tilde{\mathcal{M}}_{mnp} \coloneqq P_{[p}N_{mn]} + \frac{1}{12}(\tilde{q} \gamma_{mnp} \tilde{q}) = 0. \label{purespinortwistorconstraints3}
\end{align}

The constraints $J$ and $B$ were already considered in \cite{Berkovits:2009by} as the projective weight operator and a constraint over the physical states respectively. On the other hand, the constraint $\tilde{\mathcal{M}}_{mnp}$ corresponding to the super-Pauli-Lubanski three-form was not explicitly recognized. Interestingly, the latter constraint has already appeared in other contexts of superparticle quantization using standard spacetime variables \cite{Siegel:1987ak, Essler:1990az,Essler:1990yq}, where an infinite set of ghosts had to be introduced and the states of $D=10$ super-Yang-Mills theory were properly recovered from the corresponding BRST cohomology. It has also been remarked by Pasqua and Zumino \cite{Pasqua:2004vq} that in ten dimensions and for $\mathcal{N}=1$ supersymmetry, imposing the super-Pauli-Lubanski three-form as a constraint over the physical states -together with the more standard constraints $P^{2}=0$ and $\slashed{P} Q = 0$\,- already fixes the representation completely to be the gauge supermultiplet. In our model $P^{m} = \lambda\gamma^{m}\bar{\lambda}$ is automatically null since $\lambda^{\alpha}$ is a pure spinor, and the constraint $\slashed{P}Q$ is simply related to the $B$ constraint through $(\slashed{P} Q)^{\alpha} = 2 \lambda^{\alpha} B$. Thus, given that the constraints \eqref{purespinortwistorconstraints1}-\eqref{purespinortwistorconstraints3} appear in our description of the superparticle, it will naturally describe $D=10$ super-Yang-Mills theory. Similar remarks hold for 
the type \textrm{II}B version of \eqref{nathanvariablesin}.

The infinite set of ghosts that was introduced in \cite{Essler:1990az, Essler:1990yq} can be ultimately traced back to the fact that the constraint $\tilde{\mathcal{M}}_{mnp}$ in \eqref{purespinortwistorconstraints3} is infinitely reducible when written in a manifestly covariant fashion:
\begin{equation*}
P^{[m}\tilde{\mathcal{M}}^{npq]} = -\frac{1}{24}
(\tilde{q}\gamma^{mnpq}\lambda)B, \ \ \ \ \
(\lambda\gamma_{m})_{\alpha}\tilde{\mathcal{M}}^{mpq} = \frac{1}{3}(\lambda\gamma^{[q})_{\alpha}\bigg[(\lambda\gamma^{p]}\bar{\lambda})J - 2\Gamma^{p]}B\bigg].\label{constraintSPLintheintro}
\end{equation*}
We avoid this infinite-reducibility when uncovering the BRST operator by using the independent $SU(5)$ components of $\tilde{\mathcal{M}}^{mnp}$. This effectively truncates the ghosts-for-ghosts to two generations, and will prove particularly important in the complementary paper \cite{Sepulveda:2020wwq} where we extend the pure spinor twistor superparticle to an anomaly-free ambitwistor string \cite{Mason:2013sva}.

As usual, physical states are defined as non-trivial BRST-cohomology elements. One then finds the $D=10$ super-Yang-Mills physical states at ghost number zero, and described by the twistor superfield
\begin{align} \label{twistorsuperfieldintro}
\phi(\mu,\Gamma) = \bigg(\bar{s} & + 2\Gamma_{m}a^{m}_{-}  -4 \Gamma_{m}\Gamma_{n}s^{mn} \nonumber \\ & \hspace{0.2cm} + \frac{1}{12}(\bar{\pi}\gamma_{mnpqr}\bar{\pi})\Gamma^{m}\Gamma^{n}\Gamma^{p}h^{q}a^{r}_{+} - \frac{1}{24}(\bar{\pi}\Gamma_{mnpqr}\bar{\pi})\Gamma^{m}\Gamma^{n}\Gamma^{p}\Gamma^{q}h^{r}s\bigg)e^{\mu_{a}\bar{\pi}^{a}},
\end{align}
which carries momentum $k^{m} = \lambda\gamma^{m}\bar{\pi}$. In \eqref{twistorsuperfieldintro}, $\bar{\pi}^{a}$ with $a=1,\ldots, 5$ denotes the gauge-independent $SU(5)$ components of $\bar{\pi}^{\alpha}$, and $h^{m}$ is a constant vector satisfying $h^{m}k_{m} = 1$ whose choice does not affect \eqref{twistorsuperfieldintro}. Moreover, $a^{m} = a^{m}_{+} + a^{m}_{-}$ is the gluon polarization, with $(\bar{\pi}\gamma_{m})_{\alpha}a^{m}_{-} = (\lambda\gamma_{m})_{\alpha}a^{m}_{+} = 0$, and $\chi^{\alpha} = \bar{s}\bar{\pi}^{\alpha} + (\gamma^{mn}\lambda)^{\alpha}s_{mn} + \lambda^{\alpha}s$ is the gluino polarization, with $(\bar{\pi}\gamma^{m})_{\alpha}s_{mn} = 0$. Remarkably, this same wavefunction \eqref{twistorsuperfieldintro} was considered in \cite{Berkovits:2009by} and shown to be annihilated by the constraints \eqref{purespinortwistorconstraints1} and \eqref{purespinortwistorconstraints2}. As it turns out, it will be shown that it is additionally annihilated by the super-Pauli-Lubanski three-form with no further conditions over the twistor superfield.

%\vspace{2mm}
This work is organized as follows: In section \ref{section2} we review the Berkovits superparticle model \cite{Berkovits:1990yc}. In section \ref{section3} we define pure spinor twistors and perform a redefinition of the variables in Berkovits' model in terms of these new variables in order to obtain the new formulation of the superparticle. This procedure is highly motivated by the fact that projective pure spinors naturally realize higher-dimensional twistor transforms \cite{Hughston1,Hughston2,Hughston3,Berkovits:2004bw}. We give a brief review of these ideas in Appendix \ref{AppendixB}. Quantization of the pure spinor twistor model is performed in section \ref{section4}. We close in section \ref{section5} with some discussions and directions for further research. In Appendix \ref{AppendixA} we review some ideas of BRST quantization that are useful in sections \ref{section2} and \ref{section4}, and in Appendix \ref{AppendixB0} we outline a specific calculation illustrating usual algebraic manipulations carried out in section \ref{section3}.

\section{Review Of The Berkovits Superparticle Model} \label{section2}
In this section we review a model for the ten-dimensional massless superparticle first introduced by Berkovits in \cite{Berkovits:1990yc}. The variables in this model correspond to two ten-dimensional Majorana-Weyl bosonic spinors of opposite chirality $\Lambda^{\alpha}$, $\Omega_{\alpha}$, where $\alpha = 1,\ldots, 16$ and a fermionic vector $\psi^{m}$. In this model, the massless constraint $P^{2}=0$ is solved as $$P^{m} = (\Lambda\gamma^{m}\Lambda)$$ in virtue of the special $D=10$ gamma-matrix identity $(\gamma^{m})_{(\alpha\beta}(\gamma_{m})_{\delta)\epsilon} = 0$, where $(\gamma^{m})_{\alpha\beta}$ and $(\gamma^{m})^{\alpha\beta}$ are the ten-dimensional Pauli matrices. A set of constraints must then be imposed on the variables in order to recover the $18|8$ phase space of the Brink-Schwarz superparticle \cite{Brink:1981nb}. These constraints are given by
\begin{eqnarray}
G^{\alpha} &=& (\Lambda\gamma^{m}\Lambda)(\gamma_{m}\Omega)^{\alpha} - \Lambda^{\alpha}(\Lambda\Omega) + \psi^{m}\psi^{n}(\gamma_{m}\gamma_{n}\Lambda)^{\alpha}, \label{eq-for-Galpha}\\
T_{F} &=& (\Lambda\gamma^{m}\Lambda)\psi_{m}\label{eq-for-TF},
\end{eqnarray}
and are not independent in between them as they satisfy the reducibility relation
\begin{eqnarray}\label{eq-for-Hm}
H^{m} := (\Lambda\gamma^{m}G) - 2\psi^{m}T_{F} = 0,
\end{eqnarray}
which is itself reducible:
\begin{equation}\label{eq-for-red-Hm}
    (\Lambda \gamma^{m} \Lambda) H_{m} = 0.
\end{equation}
Counting the number of independent constraints it is straightforward to see that we indeed have eighteen bosonic and eight fermionic independent degrees of freedom.

The action principle is given by
\begin{eqnarray}\label{nathan-action}
S &=& \int d\tau \bigg(\frac{1}{2}\Lambda^{\alpha}\partial_{\tau}\Omega_{\alpha} - \frac{1}{2}\Omega_{\alpha}\partial_{\tau}\Lambda^{\alpha} - \frac{1}{2}\psi^{m}\partial_{\tau}\psi_{m} + h_{\alpha}G^{\alpha} + f T_{F}\bigg),
\end{eqnarray}
which can be readily obtained using the incidence relations connecting the standard superspace variables $(X^{m},\theta^{\alpha})$ with the $(\Lambda^{\alpha}, \Omega_{\alpha}, \psi^{m})$ variables:
\begin{equation}
\Omega_{\alpha} = (\gamma_{m}\Lambda)_{\alpha}X^{m} - \frac{1}{2}\psi^{m}(\gamma_{m}\theta)_{\alpha}, \ \ \ \psi^{m} = (\Lambda\gamma^{m}\theta)
\end{equation}
over the Brink-Schwarz superparticle action. In addition, the super-Poincar\'e generators are realized as
\begin{eqnarray}\label{sLorentzoriginal}
p_{m} = (\Lambda\gamma_{m}\Lambda), \ \ \ q_{\alpha} = (\gamma^{m}\Lambda)_{\alpha}\psi_{m}, \ \ \ M^{mn} = \frac{1}{2}(\Lambda\gamma^{mn}\Omega) + \psi^{m}\psi^{n}.
\end{eqnarray}

\subsection{Canonical Quantization}
One can now use standard first-quantization techniques and perform a quantum-mechanical analysis of the Berkovits superparticle \eqref{nathan-action}. We write the standard commutation relations for the canonical twistor variables as 
\begin{equation}
[\Lambda^{\alpha}, \Omega_{\beta}] = \delta^{\alpha}_{\beta}, \ \ \ \  \{\hat{\psi}^{m} , \hat{\psi}^{n}\} = \eta^{mn}.
\end{equation}
The $\hat{\psi}^{m}$ operators can then be represented as $SO(1,9)$ gamma-matrices $(\Gamma^{m})_{A}^{\ B}$, where $A, B = 1,\ldots, 32$, so that the ground state is described by a 32-component real spinor; namely, the reducible spinor representation of $SO(1,9)$.

Using the commutators one redefines the constraint $G^{\alpha}$ considering normal ordering contributions in such a way that the reducibility $H^{m}$ in \eqref{eq-for-Hm} still holds at the quantum level:
\begin{equation}\label{quantum-Galpha}
\hat{G}^{\alpha} = (\Lambda\gamma^{m}\Lambda)(\gamma_{m}\Omega)^{\alpha} - 2\Lambda^{\alpha}(\Lambda\Omega) + \hat{\psi}^{m}\hat{\psi}^{n}(\gamma_{m}\gamma_{n}\Lambda)^{\alpha} - 4\Lambda^{\alpha}.
\end{equation}

The quantum constraint algebra is alike the classical one and is given by
\begin{equation}\label{quantum-constraint-algebra}
[\hat{G}^{\alpha}, \hat{G}^{\beta}] = -4\Lambda^{[\alpha}\hat{G}^{\beta]}, \ \ \ [\hat{G}^{\alpha}, \hat{T}_{F}] = -2\Lambda^{\alpha}\hat{T}_{F}, \ \ \ \{\hat{T}_{F}, \hat{T}_{F}\} = 0,
\end{equation}
where $\hat{T}_{F} = (\Lambda \gamma^{m} \Lambda) \hat{\psi}_{m}$ is the quantum version of \eqref{eq-for-TF}

As usual, physical states are defined as elements annihilated by the quantum constraints. Using the irreducible $SO(1,9)$ spinor representations, the wavefunction in the $\Lambda$-representation is written as
\begin{eqnarray}
\phi_{\alpha} &=& (\Lambda\gamma^{m})_{\alpha}A_{m}\label{philower}\\
\phi^{\alpha} &=& (\Lambda B)\Lambda^{\alpha} - \frac{1}{4}(\Lambda\gamma^{m}\Lambda)(\gamma_{m}B)^{\alpha}, \label{phiupper}
\end{eqnarray}
where $A_{m}$ satisfies $(\Lambda\gamma^{m}\Lambda)A_{m} = 0$, and $A_{m}$, $B_{\alpha}$ are functions depending on $\Lambda^{\alpha}$ only through the combinations $(\Lambda\gamma^{m}\Lambda)$. One can then readily show that $\phi^{\alpha}$, $\phi_{\beta}$ are annihilated by the constraints
\begin{equation}\label{matrix-rep-GT}
(\hat{G}^{\delta})_{\alpha}{}^{\beta}\phi_{\beta} = (\hat{G}^{\delta})_{\alpha}{}^{\beta}\phi^{\alpha} = (\hat{T}_{F})^{\alpha\beta}\phi_{\beta} = (\hat{T}_{F})_{\alpha\beta}\phi^{\beta} = 0.
\end{equation}
Notice that the constraints in \eqref{matrix-rep-GT} have been written in matrix representation using a suitable identification $\hat{\psi}^{m} \rightarrow (\Gamma^{m})^{A}{}_{B}$.

Finally, the expressions in \eqref{philower}, \eqref{phiupper} can be shown to be invariant under the transformations
\begin{equation}
A_{m} \longrightarrow A_{m} + (\Lambda\gamma_{m}\Lambda)F, \ \ \ \  B_{\alpha} \longrightarrow B_{\alpha} + (\Lambda\gamma_{m}\Lambda)(\gamma^{m}F)_{\alpha},
\end{equation}
where $F$, $F^{\alpha}$ are arbitrary scalar and fermionic parameters, respectively. One then identifies $A_{m}$ with the gluon polarization and the gauge-invariant quantity $C^{\alpha} = (\Lambda\gamma^{m}\Lambda)(\gamma_{m}B)^{\alpha}$ satisfying the massless Weyl equation with the gluino polarization.

\subsection{BRST Quantization}
There exists a standard procedure to systematically construct a BRST operator for reducible gauge systems \cite{Henneaux:1992ig}. This method is based on the geometric structure satisfied by the constraint algebra on the full phase space. A brief review of this general approach can be found in Appendix \ref{AppendixA}.

As previously discussed, the Berkovits superparticle model \eqref{nathan-action} is a reducible gauge system with two levels of reducibility. Following the standard BRST quantization approach we introduce a generation of ghost variables for each reducibility level. We denote the zeroth generation of ghosts associated to the constraints $G^{\alpha}$ and $T_{F}$ by $(g_{\alpha}, f^{\alpha})$ and $(\gamma, \beta)$ respectively, the first generation of (bosonic) ghosts-for-ghosts by $(s_{m}, t^{m})$, and the second generation of (fermionic) ghosts-for-ghosts by $(\eta,\rho)$. The ghost number assignment is
\begin{align}
\#gh(g_{\alpha}) &= 1,          & & & \#gh(f^{\alpha}) &= -1,\nonumber\\
\#gh(\gamma) &= 1,          & && \#gh(\beta) &= -1,\nonumber\\
\#gh(s_{m}) &= 2,          & && \#gh(t_{m}) &= -2,\nonumber\\
\#gh(\eta) &= 3,          & &&  \#gh(\rho) &= -3.
\end{align}

The standard BRST operator for the Berkovits superparticle model can be checked to be given by
\begin{align} 
    Q =&\,g_{\alpha} :G^{\alpha}: + \gamma T_{F} + s_{m}(\Lambda \gamma^{m} f) + s_{m} (2 \psi^{m} \beta) + \eta (\Lambda \gamma_{m} \Lambda) t^{m} & \nonumber \\[0.5ex]
    & + 2 (\Lambda^{\alpha} g_{\alpha}):g_{\beta} f^{\beta}:
    - 2 (\Lambda^{\alpha} g_{\alpha}):\gamma \beta :
    -2 \big[ \Lambda^{\alpha} (\gamma_{n})_{\alpha \beta} (\gamma^{m})^{\beta \gamma} g_{\gamma} \big] :s_{m}t^{n}: & \label{BRSTop}  \\[0.5ex]
     & + 4 (\Lambda^{\alpha} g_{\alpha}):\eta \rho : + 2 \eta^{nm} s_n s_m \rho - \eta \beta^{2} - 4\Lambda^{\alpha}g_{\alpha}, \nonumber
\end{align}
where $:\,\,:$ means normally-ordered product. 

Let us delve deeper into the structure and nilpotency of the BRST operator. The first few terms correspond to the standard contributions that appear when reducibilities in the constraints are present. Notice further that all bosonic terms are linear in the ghost momenta except for the $\beta^{2}$ term, in agreement with eqn. \eqref{generalBRSToperator}. The precise coefficients are chosen to ensure nilpotency of the BRST operator. The last term can be understood as a normal ordering contribution added to have the full quantum BRST operator. Notice that adding this term is tantamount to the redefinition of $G^{\alpha}$ in \eqref{quantum-Galpha}.

The physical states are found to appear at ghost number zero, in combinations identical to the ones appearing in canonical quantization. That is,
\begin{equation} \label{V}
    V = A_{m} (\Lambda \gamma^{m})_{\beta} |0\rangle^{\beta} + \Big( (\Lambda B)\Lambda - \frac{1}{4}(\Lambda \gamma^{m} \Lambda)(\gamma_{m} B)  \Big)^{\beta} |0\rangle_{\beta} + ...
\end{equation}
where the ellipsis stands for terms at higher ghost number, is annihilated by the BRST operator $$QV=0.$$ Similarly as in canonical quantization, we identify $A_{m}$ with the gluon polarization states, and the quantity $C^{\alpha} = (\Lambda\gamma^{m}\Lambda)(\gamma_{m}B)^{\alpha}$ with the gluino polarization states, thus recovering again the $D=10$ super-Yang-Mills physical states.

\section{Pure Spinor Twistor Variables} \label{section3}

In this section we relate the twistor-like construction in \cite{Berkovits:2009by} with the ten-dimensional superparticle \cite{Brink:1981nb, Berkovits:1990yc}. This idea is strongly motivated by the results of \cite{Hughston1,Hughston2,Hughston3,Berkovits:2004bw} which establish the elegant way in which projective pure spinors realize higher-dimensional twistor transforms. A brief review of some of these ideas and results presented in \cite{Berkovits:2004bw} are provided in Appendix \ref{AppendixB}. We begin this section defining the corresponding pure spinor twistor variables after which we rewrite the superparticle model in section \ref{section2} in terms of these new variables. Then, a new twistor model containing pure spinors with an appropriate set of constraints will arise. As we shall see, many properties present in \cite{Berkovits:2009by} will naturally emerge in this new approach. Furthermore, extra constraints will be identified and they will be shown to be related to the super-Pauli-Lubanski three-form which will completely fix the physical spectrum.

\subsection{Definition Of The Pure Spinor Twistor Variables}

Pure spinor twistor variables, first considered in \cite{Berkovits:2009by}, are defined as
\begin{equation} \label{nathanvariables}
    \mathcal{Z}^{I} = (\lambda^{\alpha},\, \mu_{\alpha}, \,\Gamma^{m}), \ \ \ \ \bar{\mathcal{Z}}_{I} = (\bar{\mu}_{\alpha}, \, -\bar{\lambda}^{\alpha}, \, \bar{\Gamma}^{m}),
\end{equation}
where $\lambda^{\alpha}$ is a pure spinor, $\bar{\lambda}^{\alpha}$ is a 16-component spinor, and $\Gamma^{m}$ is a fermionic vector. The variables defining $\bar{\mathcal{Z}}_{I}$ correspond to the canonical conjugates to the variables defining $\mathcal{Z}^{I}$. By definition the variables are required to solve the constraints
\begin{align}
    S^{m} & \coloneqq (\lambda \gamma^{m} \lambda) = 0, \label{PSconstraint} \\[0.5ex]
    \label{constraintD} D & \coloneqq \lambda \mu = 0,  \\[0.5ex]
    \Phi^{mn} & \coloneqq (\lambda \gamma^{mn} \mu) + 4 \Gamma^{m} \Gamma^{n} = 0,\label{phimnconstraint} \\[0.5ex] E_{\alpha} & \coloneqq \Gamma^{m}(\lambda \gamma_{m})_{\alpha} = 0, \label{phialphaconstraint}
\end{align}
and are related to standard superspace variables through the incidence relations
\begin{equation} \label{purespinortwistorincidencerelations}
    P^{m} = (\lambda \gamma^{m} \barlambda), \ \ \ \mu_{\alpha} = (\gamma_{m} \lambda)_{\alpha} X^{m} + \Gamma^{m}(\gamma_{m} \theta)_{\alpha}, \ \ \ \Gamma_{m} = (\lambda \gamma_{m} \theta).
\end{equation}
The constraints \eqref{PSconstraint}-\eqref{phialphaconstraint} generate the following gauge transformations for the conjugate variables $\bar{\lambda}^{\alpha}$, $\bar{\mu}_{\alpha}$, $\bar{\Gamma}_{m}$:
\begin{eqnarray}\label{gaugetransformations1}
\delta \bar{\mu}_{\alpha} &=&  \mu_{\alpha}d + (\gamma^{mn}\mu)_{\alpha}\phi_{mn} + (\gamma^{m}\epsilon)_{\alpha}\Gamma_{m} + (\gamma^{m}\lambda)_{\alpha} s_{m}, \\
\delta\bar{\lambda}^{\alpha} &=& -\lambda^{\alpha}d - (\lambda\gamma^{mn})^{\alpha}\phi_{mn}, \\
\delta \bar{\Gamma}_{s} &=& 8\phi_{ms}\Gamma^{m} + (\lambda\gamma_{s}\epsilon), \label{gaugetransformations3}
\end{eqnarray}
where $s_{m}$, $d$, $\phi_{mn}$, and $\epsilon^{\alpha}$ are gauge parameters associated to the constraints \eqref{PSconstraint}-\eqref{phialphaconstraint} respectively. The linearly independent scalar bosonic currents that are invariant under \eqref{gaugetransformations1}-\eqref{gaugetransformations3} are
\begin{align}
    & \, J = \lambda \barmu - \barlambda \mu + \bar{\Gamma}^{m}\Gamma_{m}\,, \label{currentJ}\\
    & K = \lambda \barmu + \barlambda \mu \,.\label{currentK}
\end{align}
We note that $J$ is essentially the projective weight operator defined in \cite{Berkovits:2009by}. The constraints \eqref{PSconstraint}-\eqref{phialphaconstraint} induce as well a definition of non-free commutators; namely, those respecting such constraints. These commutators are given by
\begin{align}
& \left[\lambda^{\alpha}, \bar{\mu}_{\beta}\right] = \label{commutator1} \delta^{\alpha}_{\beta} - \frac{1}{2(\lambda\nu)}(\nu\gamma^{m})^{\alpha}(\gamma_{m}\lambda)_{\beta}, \\
& \left[\bar{\lambda}^{\alpha},\mu_{\beta}\right] =  \frac{1}{2(\lambda\nu)}(\nu\gamma^{m})^{\alpha}(\gamma_{m}\lambda)_{\beta}, \\
& \left[\mu_{\alpha},\bar{\mu}_{\beta}\right] =  -\frac{\nu_{\alpha}\mu_{\beta}}{(\lambda\nu)} - \frac{\mu_{\alpha}\nu_{\beta}}{(\lambda\nu)} + \frac{1}{2(\lambda\nu)}(\gamma_{m})_{\alpha\beta}(\nu\gamma^{m}\mu), \\
& \left[\Gamma_{m},\bar{\mu}_{\beta}\right] = - \frac{1}{2 (\lambda \nu)}(\gamma^{p} \gamma^{m} \nu)_{\beta}\Gamma_{p}, \\
& \{ \Gamma^{m},\bar{\Gamma}^{n} \} = \eta^{mn} - \frac{1}{2(\lambda\nu)}(\lambda\gamma^{n}\gamma^{m}\nu), \\
& \left[\mu_{\beta}, \bar{\Gamma}^{s}\right] = \frac{1}{(\lambda\nu)}(\gamma^{p}\gamma^{s}\nu)_{\beta}\Gamma_{p}, \label{commutator6}
\end{align}
where $\nu_{\alpha}$ is a fixed pure spinor $\nu \gamma^{m} \nu$ = 0. 

The super-Poincar\'e algebra can be realized with the variables \eqref{nathanvariables} through the (gauge invariant) generators
\begin{align}
    & P^{m} = (\lambda \gamma^{m} \barlambda), \\
    & N^{mn} = \frac{1}{2}(\lambda \gamma^{mn} \barmu) + \frac{1}{2}(\barlambda \gamma^{mn} \mu) - 2 \bar{\Gamma}^{[m} \Gamma^{n]},\label{definitionNps} \\
    & \tilde{q}_{\alpha} = (\barlambda \gamma^{m})_{\alpha}\Gamma_{m} - \frac{1}{2}(\lambda \gamma^{m})_{\alpha}\bar{\Gamma}_{m} \label{definitionqtildeps},
\end{align}
which are the same super-Poincar\'e generators as defined in \cite{Berkovits:2009by}. Indeed, it is straightforward to check that these representations lead to
\begin{align}
    &\{\tilde{q}_{\alpha}, \tilde{q}_{\beta} \} = \frac{1}{2}\gamma^{m}_{\alpha \beta} P_{m}, \ \ \ [\tilde{q}_{\alpha}, N_{pq}] = \frac{1}{2}\tilde{q}_{\beta}(\gamma_{pq})^{\beta}_{\ \alpha}, \ \ \ [N^{pq}, P^{s}] = 2 \eta^{s[p}P^{q]}, \\[1.0ex] & \ \ \ \ \ \ \ \ \ \ \ \ \ [N^{p q}, N^{s t}] = N^{ps} \eta^{qt} - N^{qs} \eta^{pt} - N^{pt} \eta^{qs} + N^{qt} \eta^{ps},
\end{align}
with other commutators vanishing.

\subsection{Field Redefinition And The Super-Pauli-Lubanski Constraint}

In order to relate the ten-dimensional superparticle as described in section \ref{section2} with the pure spinor twistor variables we perform a redefinition of the conjugate variables $(\Lambda^{\alpha}, \Omega_{\alpha})$ in terms of the new ones as
\begin{align}
& \Lambda^{\alpha} = \frac{\lambda^{\alpha}}{\gamma} + \frac{\gamma}{4(\lambda\nu)}(\nu\gamma_{p})^{\alpha}(\lambda\gamma_{p}\bar{\lambda}), \label{Lambda}\\ 
& \Omega_{\alpha} = \frac{2\mu_{\alpha}}{\gamma}  -\frac{\gamma}{4(\lambda\nu)}\bigg[N^{mn}(\gamma_{mn}\nu)_{\alpha} + J_{\Omega}\, \nu_{\alpha}\bigg], \label{Omega}
\end{align}
where $\gamma$ is the ghost associated to the fermionic constraint $T_{F}$ in \eqref{eq-for-TF}, $J_{\Omega}$ is the scalar current $J_{\Omega} = \lambda\bar{\mu} - 3\bar{\lambda}\mu + 2\bar{\Gamma}^{m}\Gamma_{m} = 2 J - K$, and we have further used the fixed pure spinor $\nu_{\alpha}$. Similarly, we write the fermionic variable $\psi^{m}$ as
\begin{eqnarray}
\psi^{m} &=& \frac{2\Gamma^{m}}{\gamma} + \frac{\gamma}{2}\frac{(\bar{\lambda}\gamma^{s}\gamma^{m}\nu)}{(\lambda\nu)}\Gamma_{s} - \frac{\gamma}{4}\frac{(\lambda\gamma^{s}\gamma^{m}\nu)}{(\lambda\nu)}\bar{\Gamma}_{s}. \label{psi}
\end{eqnarray}
The redefinitions \eqref{Lambda}, \eqref{Omega} and \eqref{psi} are manifestly invariant under the transformations generated by the constraints \eqref{PSconstraint}-\eqref{phialphaconstraint}. It is also readily checked that the canonical commutators are preserved under these redefinitions.

In section \ref{section2} we saw how to describe the superparticle as subjected to the constraints $G^{\alpha}$ and $T_{F}$ in equations \eqref{eq-for-Galpha} and \eqref{eq-for-TF} respectively. Under the previous redefinitions, these constraints are rewritten as
\begin{eqnarray}
T_{F} &=& \frac{\gamma}{2}\bigg[(\bar{\lambda}\gamma^{m}\bar{\lambda})\Gamma_{m} - (\lambda\gamma^{m}\bar{\lambda})\bar{\Gamma}_{m}\bigg] = \gamma B, \label{TFconstraint-with-purespinors}\\
G^{\alpha} &=& -2\Lambda^{\alpha}J - \frac{1}{4(\lambda\nu)}(\gamma_{mnp}\nu)^{\alpha}\tilde{\mathcal{M}}^{mnp}\label{Gconstraint-with-purespinors},
\end{eqnarray}
where
\begin{equation}
    \tilde{\mathcal{M}}^{pqr} = (\lambda\gamma^{[p}\bar{\lambda})N^{pq]} + \frac{1}{12}(\tilde{q}\gamma^{pqr}\tilde{q}) \label{PLconstraint}
\end{equation}
is the super-Pauli-Lubanski three-form, and where we have defined the gauge invariant quantity
\begin{equation}
    B = \frac{1}{2}\Big[(\bar{\lambda}\gamma^{m}\bar{\lambda})\Gamma_{m} - (\lambda\gamma^{m}\bar{\lambda})\bar{\Gamma}_{m}\Big]. \label{Bconstraint}
\end{equation}
This motivates the set of constraints
\begin{equation} \label{purespinortwistorconstraints}
    \{J, \,B, \, \tilde{\mathcal{M}}^{pqr}\}
\end{equation}
in the pure spinor twistor framework of the superparticle. 

Naively, one might think that \eqref{purespinortwistorconstraints} contains more independent constraints than the ones imposed by $G^{\alpha}$ and $T_{F}$ in Berkovits' model. Actually, one can show that $\tilde{\mathcal{M}}^{mnp}$ has only six independent components for a total of seven bosonic and one fermionic independent constraints. To see this, note that $\tilde{\mathcal{M}}^{mnp}$ satisfies the reducibility relations 
\begin{eqnarray}
P^{[m}\tilde{\mathcal{M}}^{npq]} &=& -\frac{1}{24}
(\tilde{q}\gamma^{mnpq}\lambda)B, \label{constraintSPL1} \\[0.5ex]
(\lambda\gamma_{m})_{\alpha}\tilde{\mathcal{M}}^{mpq} &=& \frac{1}{3}(\lambda\gamma^{[q})_{\alpha}\bigg[(\lambda\gamma^{p]}\bar{\lambda})J - 2\Gamma^{p]}B\bigg].\label{constraintSPL2}
\end{eqnarray}
     In a frame where the only non-zero component of $\lambda^{\alpha}$ is $\lambda^{+}=1$, the $SU(5)$-components of $\tilde{\mathcal{M}}^{mnp}$, namely ($\tilde{\mathcal{M}}^{abc}$, $\tilde{\mathcal{M}}^{a}{}_{bc}$, $\tilde{\mathcal{M}}^{ab}{}_{c}$, $\tilde{\mathcal{M}}_{abc}$), satisfy the relations
 \begin{align}\label{u5constraints1}
 \tilde{\mathcal{M}}_{bcd} = 0,  \ \ \ \ \ P^{[a}\tilde{\mathcal{M}}^{bcd]} = -\frac{1}{24}\epsilon^{abcde}\bar{\Gamma}_{e}B,  \\[1.0ex]
  P^{[a}\tilde{\mathcal{M}}^{b]}{}_{cd} = 0, \ \ \ \ \  P^{[a}\tilde{\mathcal{M}}^{bc]}{}_{d} = -\frac{1}{16}\delta^{[a}_{d}\bar{\lambda}^{b}\Gamma^{c]}B, \label{u5constraints2}
 \end{align}
 where $a,b,\ldots = 1,\ldots, 5$. The second of \eqref{u5constraints1} and the two equations in \eqref{u5constraints2} are not independent as one can antisymmetrize them with $P^{a}$ to find they identically vanish. Likewise, the tensors defined through this antisymmetrization are not independent and the antisymmetrization of them with the momentum $P^{a}$ also vanishes. This antisymmetrization procedure defines a chain of constraints that eventually finishes when there are ten indices antisymmetrized. One then finds various reducibility relations for the $SU(5)$-components of $\tilde{\mathcal{M}}^{mnp}$, and one is left with 6 components for $\tilde{\mathcal{M}}^{abc}$, 20 components for $\tilde{\mathcal{M}}^{ab}{}_{c} $, and 10 components for $\tilde{\mathcal{M}}^{a}{}_{bc}$. Moreover, equation \eqref{constraintSPL2} allows one to show that $\tilde{\mathcal{M}}^{a}{}_{bc}$, $\tilde{\mathcal{M}}^{ab}{}_{c}$ are actually related to the constraints $J$, $B$. Therefore, the only independent components of $\tilde{\mathcal{M}}^{mnp}$ are given by the six independent components of $\tilde{\mathcal{M}}^{abc}$. 
 
Finally, when we rewrite the super-Lorentz currents of the Berkovits' model in terms of the pure spinor twistor variables, we obtain
\begin{align}
q_{\alpha} &= \tilde{q}_{\alpha} + \frac{\nu_{\alpha} \gamma^{2}}{2 (\lambda \nu)}B, \label{definitionq} \\
M^{mn} &= N^{mn} - \frac{\gamma^{2}}{32 (\lambda \nu)^{2}} (\nu \gamma^{mn} \gamma^{pqr} \nu) \tilde{\mathcal{M}}_{pqr}\label{definitionMmn}.
\end{align}
An explicit demonstration of these relations can be found in Appendix \ref{AppendixB0}. Thus, we see that the super-Lorentz currents of both models are related to each other up to unphysical terms that give vanishing contributions in the algebra. 

\section{The Pure Spinor Twistor Superparticle} \label{section4}
Considering the maps \eqref{Lambda}, \eqref{Omega}, \eqref{psi} defined in the previous section, one defines the pure spinor twistor superparticle action to be
\begin{equation} \label{purespinortwistormodel}
    S = \int d\tau \left[\bar{\mathcal{Z}}_{I}\partial_{\tau}\mathcal{Z}^{I} + \epsilon J + \chi B + \Upsilon_{mnp}\tilde{\mathcal{M}}^{mnp} \right],
\end{equation}
where the algebra formed by the first-class constraints $J$, $B$, $\tilde{\mathcal{M}}^{mnp}$ is given by
\begin{align}
& [J, J] = 0, \ \ \ [J, B] = B, \ \ \  \{B,B\} = 0, \ \ \     
[J,\tilde{\mathcal{M}}^{mnp}] = 0, \ \ \ [B, \tilde{\mathcal{M}}^{mnp}] = 0, \nonumber \\[1.0ex]
 & \ \ \ \ \ \ \ \ \ \ \ \ \ \ \ \  [\tilde{\mathcal{M}}_{pqr}, \tilde{\mathcal{M}}^{stu}] =    6 (\lambda \gamma_{[r} \barlambda) \delta^{[s}_{p} \tilde{\mathcal{M}}_{q]}^{t u]} - \frac{1}{36} (\tilde{q} \gamma_{pqr} \gamma ^{stu} \lambda) B.
 \end{align}

The full set of gauge transformations for the supertwistor variables is given by
\begin{align}
    \delta \lambda^{\alpha} & = \, j \lambda^{\alpha} + \frac{m_{pqr}}{2}P^{[r}(\gamma^{pq]}\lambda)^{\alpha}, \\
    \delta \mu_{\alpha} & = \, j \mu_{\alpha} + \kappa \tilde{q}_{\alpha}+ m_{pqr}\bigg[ (\lambda \gamma^{[r})_{\alpha}N^{pq]} + \frac{1}{2}P^{[r}(\gamma^{pq]}\mu)_{\alpha} + \frac{1}{6}(\tilde{q} \gamma^{pqr} \gamma^{m})_{\alpha}\Gamma_{\alpha} \bigg], \\
    \delta \Gamma^{m} & = \, j \Gamma^{m} -\frac{1}{2}\kappa P^{m} +m_{pqr} \bigg[ 2 P^{\bar{p}} \eta^{m \bar{p}} \Gamma^{\bar{q}} - \frac{1}{12} (\tilde{q} \gamma^{pqr} \gamma^{m} \lambda) \bigg], \\ \delta\bar{\lambda}^{\alpha} & = - j\barlambda^{\alpha} - \lambda^{\alpha}d - (\lambda\gamma^{mn})^{\alpha}\phi_{mn} + \frac{m_{pqr}}{2}P^{[r}(\gamma^{pq]}\barlambda)^{\alpha },\\[1.5ex] \delta \barmu_{\alpha} &  = - j \barmu_{\alpha} -\frac{\kappa}{2} (\barlambda \gamma^{m})_{\alpha}\bar{\Gamma}_{m} + \mu_{\alpha}d + (\gamma^{mn}\mu)_{\alpha}\phi_{mn} + (\gamma^{m}\lambda)_{\alpha}s_{m} + (\gamma^{m}\epsilon)_{\alpha}\Gamma_{m} \nonumber \\  & \hspace{0.68cm} + m_{pqr}\bigg[ (\barlambda \gamma^{[r})_{\alpha}N^{pq]} + \frac{1}{2}P^{[r}(\gamma^{pq]} \barmu)_{\alpha} - \frac{1}{12}(\tilde{q} \gamma^{pqr} \gamma^{m})_{\alpha} \bar{\Gamma}_{m} \bigg], \ \\[1.0ex]
    \delta \bar{\Gamma}_{m} & = - j \bar{\Gamma}_{m}+\frac{\kappa}{2}(\barlambda \gamma_{m} \barlambda) +  8\phi_{pm}\Gamma^{p} + (\lambda\gamma_{m}\epsilon) + m_{pqr}\bigg[ 2P^{\bar{r}}\delta_{m}^{\bar{p}} \bar{\Gamma}^{\bar{q}} + \frac{1}{6}(\tilde{q} \gamma^{pqr} \gamma_{m} \barlambda) \bigg],
\end{align}
where $s_{m}$, $d$, $\phi_{mn}$, $\epsilon^{\alpha}$ are gauge parameters associated to the constraints \eqref{PSconstraint}-\eqref{phialphaconstraint} and $j$, $\kappa$ and $m_{pqr}$ are gauge parameters associated to \eqref{currentJ}, \eqref{Bconstraint} and \eqref{PLconstraint} respectively.

\vspace{2mm}

\subsection{Canonical Quantization}
Following the standard procedure one promotes the first-class constraints to be operators acting on the Hilbert space and annihilating physical states. The wavefunction $\phi(Z)$ must then satisfy
\begin{equation}  \label{quantumconstraints}
(J+1)(\phi(Z)) = 0, \ \ \
B(\phi(Z)) = 0, \ \ \ 
\tilde{\mathcal{M}}^{mnp}(\phi(Z)) = 0,
\end{equation}
where the identity operator in the first equation comes from normal ordering ambiguities present in the reducibility relation \eqref{constraintSPL2}. The corresponding field annihilated by these constraints is given by the projective weight -1 wavefunction
\begin{eqnarray}\label{wavefunction}
\phi(Z) = \Phi(\Gamma)e^{\mu_{a}\bar{\pi}^{a}},
\end{eqnarray}
where
\begin{align} \label{twistorsuperfield}
\Phi(\Gamma) = \bar{s} + 2\Gamma_{m}a^{m}_{-} & - 4 \Gamma_{m}\Gamma_{n}s^{mn} \nonumber \\ &+ \frac{1}{12}(\bar{\pi}\gamma_{mnpqr}\bar{\pi})\Gamma^{m}\Gamma^{n}\Gamma^{p}h^{q}a^{r}_{+} - \frac{1}{24}(\bar{\pi}\Gamma_{mnpqr}\bar{\pi})\Gamma^{m}\Gamma^{n}\Gamma^{p}\Gamma^{q}h^{r}s.
\end{align}
In \eqref{wavefunction}, we have used the pureness of $\lambda^{\alpha}$ to set eleven components of $\bar{\pi}^{\alpha}$ to zero, so that one is left with the 5-component vector $\bar{\pi}^{a}$, $a=1,\ldots, 5$ transforming in the fundamental of $SU(5)$. Furthermore, $a^{m}$ in \eqref{twistorsuperfield} denotes the gluon polarization, $a^{m} = a^{m}_{+} + a^{m}_{-}$ with $(\bar{\pi}\gamma_{m})_{\alpha}a^{m}_{-} = (\lambda\gamma_{m})_{\alpha}a^{m}_{+} = 0$, the gluino polarization has been written as $\chi^{\alpha} = \bar{s}\bar{\pi}^{\alpha} + (\gamma^{mn}\lambda)^{\alpha}s_{mn} + \lambda^{\alpha}s$, with $(\bar{\pi}\gamma^{m})_{\alpha}s_{mn} = 0$, and $h^{m}$ is a constant vector satisfying $h^{m}k_{m} = 1$ whose choice will not affect \eqref{twistorsuperfield} since $\bar{\pi}^{a}$ automatically satisfies $\bar{\pi}\gamma^{m}\bar{\pi} = 0$. Note that under the transformation $\lambda^{\alpha} \rightarrow t^{-1} \lambda^{\alpha}$, $\bar{\pi}^{\alpha} \rightarrow t \bar{\pi}^{\alpha}$, the physical components appearing in \eqref{twistorsuperfield} scale as
\begin{equation}
a_{-} \rightarrow a_{-}, \ \ \ a_{+}\rightarrow a_{+}, \ \ \ \bar{s}\rightarrow t^{-1}\bar{s}, \ \ \ s_{mn}\rightarrow s_{mn}, \ \ \ s\rightarrow ts.
\end{equation}

Remarkably, this very same wavefunction has already been considered in \cite{Berkovits:2009by}, where the $D=10$ super-Yang-Mills physical fields were shown to appear as $\phi(Z) = \Phi(\Gamma)e^{\mu \bar{\pi}}$ carrying momentum $k^{m} = \lambda\gamma^{m}\bar{\pi}$, with $\lambda^{\alpha}$ a projective pure spinor. %stands for the eigenvalue of $\barlambda^{\alpha}$.
Furthermore, a constraint
\begin{equation}\label{constraintB}
B(\phi(Z)) = 0,
\end{equation}
with $$B = (\bar{\lambda}\gamma^{m}\bar{\lambda})\Gamma_{m} - (\lambda\gamma^{m}\bar{\lambda})\bar{\Gamma}_{m}$$ had to be imposed. This is effectively the same $B$ constraint considered in this work, but that we found naturally in our construction of the pure spinor twistor superparticle.

 As in \cite{Berkovits:2009by}, the wavefunction \eqref{wavefunction} satisfies the first two physical state conditions in \eqref{quantumconstraints}. It turns out it also satisfies the third requirement $\tilde{\mathcal{M}}^{mnp}(\phi(Z)) = 0$. This can be seen as follows: In the frame where the only non-zero component of $\lambda^{\alpha}$ is $\lambda^{+} = 1$, one can choose a gauge where the only non-zero $SU(5)$ components of $\bar{\mu}_{\alpha}$, $\bar{\lambda}^{\alpha}$, $\bar{\Gamma}^{m}$ are 
 \begin{equation}
 (\bar{\mu}_{+}, \, \bar{\mu}^{ab}), \ \ \ \bar{\lambda}^{a}, \ \ \ \bar{\Gamma}_{b}, \ \ \  \mbox{$a,b = 1,\ldots, 5$},
 \end{equation}
 transforming in the (singlet, ten-dimensional antisymmetric), fundamental, antifundamental of $SU(5)$, respectively. In addition, after solving the constraints \eqref{constraintD}-\eqref{phialphaconstraint}, one finds that the only non-zero $SU(5)$ components of $\mu_{\alpha}$, $\Gamma^{m}$ are $\mu_{j}$, $\Gamma^{j}$, respectively. Therefore, the only independent components of $\tilde{\mathcal{M}}^{mnp}$, namely $\tilde{\mathcal{M}}^{abc}$, take the form
\begin{eqnarray}
\tilde{\mathcal{M}}^{abc} &=& \bar{\lambda}^{[a}\bar{\mu}^{bc]} + \frac{1}{12}\epsilon^{abcde}\bar{\Gamma}_{d}\bar{\Gamma}_{e}.
\end{eqnarray}
One can now let $\tilde{\mathcal{M}}^{abc}$ act on the gluon sector of $\phi(Z)$ in \eqref{twistorsuperfield}. The contribution coming from the term proportional to $a_{-}^{m}$ exactly cancels the contribution due to the term proportional to $a_{+}^{m}$. A similar argument follows for the fermionic sector. Hence, we have recovered the twistor superfield $\phi(Z)$ of \cite{Berkovits:2009by} as the field satisfying the physical state conditions \eqref{quantumconstraints} of the pure spinor twistor model. This is not surprising, since the vanishing of the super-Pauli-Lubanski tensor actually completely fixes the supersymmetry representation in any spacetime dimension as discussed in \cite{Pasqua:2004vq}. Actually, the $D=10$ super-Yang-Mills multiplet encoded by the twistor superfield $\phi(Z)$ must necessarily satisfy this requirement, since the twistor superfield $\phi(Z) = \Phi(\Gamma)e^{\mu\bar{\pi}}$ is exactly the same as the unintegrated vertex operator of the standard pure spinor formalism $U = \lambda^{\alpha}A_{\alpha}$ after using the incidence relations \eqref{purespinortwistorincidencerelations}. On the other hand, one might wonder if a similar result regarding the super-Pauli-Lubanski three-from is valid in the standard pure spinor formalism. The answer turns out to be affirmative. More precisely, one can show that the super-Pauli-Lubanski three-form annihilates $U=\lambda^{\alpha}A_{\alpha}$, up to BRST-exact terms. 

\subsection{BRST Quantization}
We find convenient for BRST quantization to consider only the independent components of the super-Pauli-Lubanski constraint. As discussed before, these components are given by
\begin{equation}
\hspace{2cm} \tilde{\mathcal{M}}^{abc} = P^{[a}N^{bc]} + \frac{1}{12}(\tilde{q}\gamma^{abc}\tilde{q}), \hspace{0.5cm} a,b,c = 1, \ldots, 5,
\end{equation}
and the only non-trivial commutator in the algebra generated by $(J,B,\tilde{\mathcal{M}}^{abc})$ is:
\begin{equation}
    [J,B] = B,
\end{equation}
with other (anti)commutators vanishing.

The non-Lorentz covariant constraint $\tilde{\mathcal{M}}^{abc}$ is not irreducible. In fact, one can show that \pagebreak
\begin{eqnarray}
H^{abcd} &:= & P^{[a}\tilde{\mathcal{M}}^{bcd]} + \frac{1}{24}\epsilon^{abcde}\tilde{q}_{e}\lambda^{+}B = 0, \label{psred1}\\
L^{abcde} &:= & P^{[a}H^{bcde]} = 0, \label{psred2}
\end{eqnarray}
where $\tilde{q}_{e}$ is the $SU(5)$ antifundamental vector of $\tilde{q}_{\alpha}$, and $\lambda^{+}$ is the $SU(5)$ scalar of the pure spinor $\lambda^{\alpha}$.

Following the standard prescription for BRST quantization, we introduce a zeroth generation of conjugate ghost variables $(\sigma, \tilde{\sigma})$, $(\gamma, \beta)$, $(f_{abc}, \tilde{f}^{abc})$ for the constraints $(J, B, \tilde{\mathcal{M}}^{abc})$ respectively.  The reducibility \eqref{psred1} will give rise to the introduction of a first generation of (bosonic) ghosts-for-ghosts $(s_{abcd}, \tilde{s}^{abcd})$, and \eqref{psred2} will imply the presence of a second generation of (fermionic) ghosts-for-ghosts $(f^{abcde},\tilde{f}_{abcde})$. The (anti)commutators for the ghosts are taken as
\begin{equation}
     \{\tilde{f}^{abc}, f_{def} \} = \delta^{abc}_{def}, \ \ \ [s_{abcd}, \tilde{s}^{efgh}] = \delta^{efgh}_{abcd}, \ \ \  \{\tilde{f}^{abcde}, f_{fghij} \} = \delta^{abcde}_{fghij},
\end{equation}
where $\delta^{\mu_{1}, \ldots, \mu_{p}}_{\nu_{1}, \ldots, \nu_{p}} := \delta^{\mu_{1}}_{[\nu_{1}} \ldots \delta^{\mu_{p}}_{\nu_{p}]}$. All ghosts are antisymmetric in their indices. Then, the BRST operator can be checked to be given by
\begin{align}\label{psQ}
Q = \, \sigma J + \gamma B + f_{abc}\tilde{\mathcal{M}}^{abc} + \sigma \gamma\beta + s_{abcd} \Big[ \tilde{f}^{abc}P^{d} + \frac{1}{4!}\epsilon^{abcde}\tilde{q}_{e}&\lambda^{+}\beta \Big] \nonumber \\[0.5ex]
+ f_{abcde}P^{a}\tilde{s}^{bcde} + \frac{1}{5!}(\lambda^{+})^{2}\epsilon^{abcde}f_{abcde}\beta^{2} + \sigma. &
\end{align}
The terms in the first line of \eqref{psQ} arise from the constraint algebra and the reducibility \eqref{psred1}. The second line takes into account the reducibility \eqref{psred2}. The nilpotency of \eqref{psQ} readily follows from the fact that $\tilde{\mathcal{M}}^{abc}$ has vanishing commutators with all coefficients of \eqref{psred1}, \eqref{psred2}.

Physical states are then found at ghost number zero cohomology and described by the same twistor superfield \eqref{wavefunction}.

\section{Discussion and Future Directions} \label{section5}

In this work we have developed a model of the superparticle that is clearly intertwined with the twistor-like construction of \cite{Berkovits:2009by}. The variables $\mathcal{Z}^{I}=(\lambda^{\alpha},\, \mu_{\alpha},\, \Gamma^{m}), \ \bar{\mathcal{Z}}_{I}=(\barmu_{\alpha},\, -\barlambda^{\alpha},\, \bar{\Gamma}^{m})$ naturally appear in this formulation, and we have rediscovered the constraints $J$ and $B$ already present in \cite{Berkovits:2009by}. Additionally, we have found that our formulation presents a constraint
\begin{equation} \label{paulu}
    \tilde{\mathcal{M}}_{mnp} := P_{[p}N_{mn]} + \frac{1}{12}(\tilde{q} \gamma_{mnp} \tilde{q}) = 0,
\end{equation}
which was not recognized in \cite{Berkovits:2009by} and that corresponds to the super-Pauli-Lubanski three-form. Furthermore, we found that the superfield considered in \cite{Berkovits:2009by} is not only annihilated by $B$, but by the constraint \eqref{paulu} as well. We found the model, and specially the associated constraints, by relating a set of variables from another model of the superparticle \cite{Berkovits:1990yc} to the pure spinor variables used here, and studied quantization of both models. \pagebreak

Many further directions of research arise from this work. First and foremost, one may straightforwardly generalize the construction presented here to its type IIB version through a simple extension of the supertwistors \eqref{nathanvariables} including an additional set of constrained fermionic variables. More interestingly, one might try to construct a prescription for computing scattering amplitudes using ideas inspired in light-cone gauge functional integration, as done in \cite{Berkovits:2019bbx} for the Berkovits superparticle model studied in section \ref{section2}. This will require the fixing of all the gauge symmetries generated by $(J, B, \tilde{M}^{abc})$ and the introduction of interaction-point and physical operators properly regularized. It would be interesting to see how these operators actually look like and how the equivalence between the gauge-fixed pure spinor and light-cone RNS ambitwistor models is realized. On the other hand, one can also promote the superparticle model presented here to a worldsheet action in order to possibly recover the amplitudes construction presented in \cite{Berkovits:2009by}. A specific instance of this idea consists in promoting the pure spinor twistor superparticle action into an ambitwistor string action \cite{Mason:2013sva}, which is worked out in the complementary paper \cite{Sepulveda:2020wwq}. It would be very interesting to study if further worldsheet actions, perhaps with non-zero string tension, may be constructed using the ideas presented in this work. In \cite{Berkovits:1990yr}, for instance, the ideas presented in the Berkovits' superparticle model \cite{Berkovits:1990yc} were applied to the superstring. Similar ideas may be applied using the pure spinor twistor variables as a basis.

A different route for further research would be to work out similar constructions for the $D=11$ superparticle in terms of matter pure spinor variables and imposing, as a definition of the model, the super-Pauli-Lubanski three-form as a constraint. Following Pasqua and Zumino \cite{Pasqua:2004vq}, imposing this constraint would fix the spectrum to be the eleven-dimensional supergravity multiplet. Similarly, one might try to find the Type IIA extension of the $D=10$ model here presented. Presumably, this will require a twistor transform which considers opposite chiralities instead of a single chirality.

Although we found the physical states in the BRST quantization of the Berkovits' superparticle model at ghost number zero in \eqref{V}, it would be interesting to study further ghost number sectors and see if extra fields appear in the BRST-closed states. This will probably require to consider the action of the non-scalar ghosts over their respective ground states.

For the BRST quantization of the pure spinor twistor superparticle, we expressed the BRST operator in terms of the independent components $\tilde{\mathcal{M}}^{abc}$ of the super-Pauli-Lubanski three-form. This will prove to be particularly useful for the construction of the corresponding anomaly-free ambitwistor string in \cite{Sepulveda:2020wwq}, although it would be interesting to study BRST quantization considering the full $\tilde{\mathcal{M}}^{mnp}$ constraint in the BRST operator.

\acknowledgments
We would like to thank Renann Lipinski Jusinskas for useful discussions. We are also grateful to Nathan Berkovits and Oliver Schlotterer for enlightening discussions and reading the manuscript. D.G.S would like to thank the Abdus Salam International Centre for Theoretical Physics, ICTP-SAIFR/IFT- UNESP, FAPESP grant 2016/01343-7, CAPES-PROEX, and Perimeter Institute for partial financial support. M.G. was supported by the European Research Council under ERC-STG-804286 UNISCAMP. This research was supported in part by Perimeter Institute for Theoretical Physics. Research at Perimeter Institute is supported by the Government of Canada through the Department of Innovation, Science, and Economic Development, and by the Province of Ontario through the Ministry of Research and Innovation.

\appendix

\section{BRST Quantization for Reducible Gauge Systems} \label{AppendixA}
We will briefly review the ideas developed in \cite{Henneaux:1992ig} for quantizing systems possessing reducible gauge symmetries. Let $G_{a_{0}}$ be a constraint of the $L$-reducible gauge system satisfying $[G_{a_{0}}, G_{b_{0}}] = f_{a_{0}b_{0}}{}^{c_{0}}G_{c_{0}}$ where $f_{a_{0}b_{0}}{}^{c_{0}}$ is, in general, a function of the matter variables. With no loss of generality, $G_{a_{0}}$ will be assumed to be bosonic. The general case easily follows from this by grading commutators where needed. Let us denote the reducibility functions of level $k$ by $Z_{a_{k}}{}^{a_{k-1}}$, therefore
\begin{eqnarray}
Z_{a_{1}}{}^{a_{0}}G_{a_{0}} &=& 0,\label{reducibility-chain1}\\ Z_{a_{2}}{}^{a_{1}}Z_{a_{1}}{}^{a_{0}} &=& f_{a_{2}}{}^{a_{0}b_{0}}G_{b_{0}},\label{reducibility-chain2}\\
\vdots\nonumber\\
Z_{a_{L}}{}^{a_{L-1}}Z_{a_{L-1}}{}^{a_{{L-2}}} &=& f_{a_{L}}{}^{a_{L-2} b_{0}}G_{b_{0}}.\label{reducibility-chain3}
\end{eqnarray}

Before constructing the BRST charge, it will be useful to first discuss a set of identities coming from the reducibility structure \eqref{reducibility-chain1}-\eqref{reducibility-chain3}. One then starts with the Jacobi identity
\begin{eqnarray}
[[G_{[a_{0}},G_{b_{0}}],G_{c_{0}]}] &=& 
\left[-f_{a_{0}b_{0}}{}^{d_{0}}f_{c_{0}d_{0}}{}^{e_{0}} + \partial_{c_{0}}f_{a_{0}b_{0}}{}^{e_{0}}\right]G_{e_{0}},
\end{eqnarray}
where $U_{a_{0}b_{0}c_{0}}{}^{a_{1}}$ is a function of matter variables, completely antisymmetric in $a_{0}, b_{0}, c_{0}$, and we used the notation $\partial_{a_{0}}F = [F, G_{a_{0}}]$. One then learns that
\begin{eqnarray}\label{identity-U}
-f_{[a_{0}b_{0}}{}^{d_{0}}f_{c_{0}]d_{0}}{}^{e_{0}} + \partial_{[c_{0}}f_{a_{0}b_{0}]}{}^{e_{0}} &=& -\frac{2}{3}U_{a_{0}b_{0}c_{0}}{}^{a_{1}}Z_{a_{1}}{}^{e_{0}},
\end{eqnarray}
where the overall factor $-\frac{2}{3}$ was chosen for convenience. In addition, one can take a derivative $\partial_{b_{0}}$ on both sides of eqn. \eqref{reducibility-chain1} to obtain
\begin{eqnarray}
\left[Z_{a_{1}}{}^{a_{0}}f_{b_{0}a_{0}}{}^{d_{0}} - \partial_{b_{0}}Z_{a_{1}}{}^{d_{0}}\right]G_{d_{0}} &=& 0,
\end{eqnarray}
and so
\begin{eqnarray}\label{identity-D}
Z_{a_{1}}{}^{a_{0}}f_{b_{0}a_{0}}{}^{d_{0}} - \partial_{b_{0}}Z_{a_{1}}{}^{d_{0}} &=& -D_{a_{1}b_{0}}{}^{b_{1}}Z_{b_{1}}{}^{d_{0}}.
\end{eqnarray}
After taking a further derivative on eqn. \eqref{identity-U} and using eqns. \eqref{reducibility-chain1}, \eqref{reducibility-chain2} one finds that
\begin{eqnarray}\label{ma0b0c0d0a2}
\frac{2}{3}\partial_{[d_{0}}U_{a_{0}b_{0}c_{0}]}{}^{b_{1}} + \frac{2}{3}U_{[a_{0}b_{0}c_{0}}{}^{a_{1}}D_{a_{1}d_{0}]}{}^{b_{1}} + f_{[a_{0}b_{0}}{}^{g_{0}}U_{c_{0}g_{0}d_{0}]}{}^{b_{1}} = M_{a_{0}b_{0}c_{0}d_{0}}{}^{a_{2}}Z_{a_{2}}{}^{a_{1}},
\end{eqnarray}
where $M_{a_{0}b_{0}c_{0}d_{0}}{}^{a_{2}}$ is a function of matter variables, completely antisymmetric in $a_{0}, b_{0}, c_{0}, d_{0}$, and it is subjected to the consistency condition
\begin{eqnarray}\label{consistency-for-ma0b0c0d0a2}
M_{a_{0}b_{0}c_{0}d_{0}}{}^{a_{2}}f_{a_{2}}{}^{e_{0}g_{0}} &=& \frac{1}{2}[f_{[a_{0}b_{0}}{}^{e_{0}},f_{d_{0}c_{0}]}{}^{g_{0}}].
\end{eqnarray}
On the other hand, a further derivation on both sides of eqn. \eqref{identity-D} and the use of eqns. \eqref{reducibility-chain1}, \eqref{reducibility-chain2}, \eqref{identity-U} provide us the identity 
\begin{equation}\label{sa0b0a1a2}
\partial_{[e_{0}}D_{a_{1}b_{0}]}{}^{b_{1}} - Z_{a_{1}}{}^{a_{0}}U_{e_{0}b_{0}a_{0}}{}^{b_{1}} - \frac{1}{2}f_{b_{0}e_{0}}{}^{h_{0}}D_{a_{1}h_{0}}{}^{b_{1}} + D_{a_{1}[b_{0}}{}^{b_{1}}D_{b_{1}e_{0}]}{}^{b_{1}} = S_{e_{0}b_{0}a_{1}}{}^{a_{2}}Z_{a_{2}}{}^{b_{1}},
\end{equation}
where $S_{e_{0}b_{0}a_{1}}{}^{a_{2}}$ is a function of matter variables, completely antisymmetric in $e_{0}, b_{0}$, and it is subjected to the consistency relation
\begin{eqnarray}\label{consistency-for-sa0b0a1a2}
S_{e_{0}b_{0}a_{1}}{}^{a_{2}}f_{a_{2}}{}^{f_{0}h_{0}} &=& \frac{1}{2}[f_{b_{0}e_{0}}{}^{f_{0}}, Z_{a_{1}}{}^{h_{0}}] - \frac{1}{2}[f_{b_{0}e_{0}}{}^{h_{0}}, Z_{a_{1}}{}^{f_{0}}].
\end{eqnarray}
Moreover, a similar derivation procedure applied to eqn. \eqref{reducibility-chain2} gives us the identity
\begin{eqnarray}\label{na0a2b2}
\partial_{e_{0}}Z_{a_{2}}{}^{a_{1}} + Z_{a_{2}}{}^{f_{1}}D_{f_{1}e_{0}}{}^{a_{1}} &=& N_{e_{0}a_{2}}{}^{b_{2}}Z_{b_{2}}{}^{a_{1}},
\end{eqnarray}
where $N_{e_{0}a_{2}}{}^{b_{2}}$ is a function of matter variables and satisfies the consistency equation
\begin{eqnarray}\label{consistency-for-na0a2b2}
N_{e_{0}a_{2}}{}^{b_{2}}f_{b_{2}}{}^{a_{0}b_{0}} &=& \partial_{e_{0}}f_{a_{2}}{}^{a_{0}b_{0}} + f_{a_{2}}{}^{a_{0}g_{0}}f_{g_{0}e_{0}}{}^{b_{0}} - f_{a_{2}}{}^{g_{0}b_{0}}f_{e_{0}g_{0}}{}^{a_{0}}.
\end{eqnarray}
One more identity can be found by multiplying by $Z_{a_{1}}{}^{b_{0}}$ on both sides of eqn. \eqref{identity-D} and symmetrizing the first level reducibility indices 
\begin{eqnarray}\label{ra1b1a2}
Z_{(d_{1}}{}^{b_{0}}D_{a_{1})b_{0}}{}^{b_{1}} &=& R_{d_{1}a_{1}}{}^{a_{2}}Z_{a_{2}}{}^{b_{1}},
\end{eqnarray}
where $R_{d_{1}a_{1}}{}^{a_{2}}$ is a function of matter variables, completely symmetric in $a_{1}, b_{1}$, and satisfies
\begin{eqnarray}\label{consistency-for-ra1b1a2}
R_{d_{1}a_{1}}{}^{a_{2}}f_{a_{2}}{}^{b_{0}d_{0}} &=& [Z_{(d_{1}}{}^{b_{0}}, Z_{a_{1})}{}^{d_{0}}].
\end{eqnarray}

We will stop here since these identities are the only relevant ones for the construction of the BRST charge of the 2-reducible supertwistor model \eqref{nathan-action}, as we shall see below. However, the procedure above illustrated can easily be continued to find a complete set of identities satisfied by the reducibility functions $Z_{a_{k}}{}^{a_{k-1}}$, $f_{a_{k}}{}^{a_{k-2} a_{0}}$ and the structure coefficients $f_{a_{0}b_{0}}{}^{c_{0}}$ of a general $L$-reducible model.

To construct the BRST charge, one first introduces a couple of conjugate ghost variables $(\eta^{a_{k}}, \mathcal{P}_{k})$ for each reducibility relation of the form \eqref{reducibility-chain1}-\eqref{reducibility-chain3}. The assignment of ghost number is
\begin{eqnarray}
\#gh(\eta^{a_{k}}) = k+1, \ \ \ \#gh (\mathcal{P}_{a_{k}}) = -k-1, \hspace{4mm} k = 0,1,\ldots, L,
\end{eqnarray}
and the antighost number is
\begin{eqnarray}
\#antigh(\eta^{a_{k}}) = 0, \ \ \ \#antigh (\mathcal{P}_{a_{k}}) = k+1, \hspace{4mm} k = 0,1,\ldots, L.
\end{eqnarray}
One then expands the BRST operator $Q$ in terms of functions of fixed antighost number
\begin{eqnarray}
Q &=& \sum_{p=0}^{N}\Omega^{(p)},
\end{eqnarray}
where the antighost number of $\Omega^{(p)}$ is $p$. The nilpotency property of $Q$ then requires 
\begin{eqnarray}\label{nilpotency-condition}
\{Q, Q\} &=& \sum_{p=0}^{N}B^{(p)} = 0,
\end{eqnarray}
where $B^{(p)}$ possesses antighost number $p$ and takes the form
\begin{eqnarray}\label{expression-for-B}
B^{(p)} &=& \sum_{k=0}^{p}\{\Omega^{(p-k)}, \Omega^{(k)}\}_{o} + \sum_{k=0}^{p+1}\sum_{s=0}^{k}\{\Omega^{(p-k+s+1}, \Omega^{(k)}\}_{\eta^{a_s},\mathcal{P}_{a_s}}.
\end{eqnarray}
Therefore one learns from eqn. \eqref{nilpotency-condition} that 
\begin{eqnarray}
B^{(p)} = 0 \hspace{2mm} \mbox{for $p = 0, 1, 2,\ldots$}
\end{eqnarray}
One can now use the recursive formula \eqref{nilpotency-condition} and find $\Omega^{(p)}$ for each $p=0,1,\ldots, L$. There is only one object constructed out of ghost fields and reducibility functions carrying ghost number 1 and antighost number 0, namely
\begin{eqnarray}\label{omega0}
\Omega^{(0)} &=& \eta^{a_{0}}G_{a_{0}}.
\end{eqnarray}
After plugging \eqref{omega0} into \eqref{expression-for-B} for $p=0$, one finds
\begin{eqnarray}
B^{(0)} &=& f_{a_0 b_0}{}^{c_0}\eta^{a_0}\eta^{b_0}G_{c_0} + 2\frac{\partial \Omega^{(1)}}{\partial\mathcal{P}_{a_{0}}}G_{a_{0}},
\end{eqnarray}
which leads us to the formula
\begin{eqnarray}\label{omega1}
\Omega^{(1)} &=& Z_{a_{1}}{}^{a_{0}}\eta^{a_{1}}\mathcal{P}_{a_{0}} - \frac{1}{2}f_{a_0 b_0}{}^{c_0}\eta^{a_0}\eta^{b_0}\mathcal{P}_{c_0}.
\end{eqnarray}
One can now replace eqns. \eqref{omega0}, \eqref{omega1} in \eqref{expression-for-B} for $p=1$ to find
\begin{eqnarray}
B^{(1)} &=& [\partial_{e_0}f_{a_0 b_0}{}^{c_0} - f_{d_{0}b_{0}}{}^{c_{0}}f_{a_{0}e_{0}}{}^{d_{0}}]\eta^{e_{0}}\eta^{a_{0}}\eta^{b_{0}}\mathcal{P}_{c_{0}} - 2[\partial_{b_{0}}Z_{a_{1}}{}^{a_{0}} + f_{d_{0}b_{0}}{}^{a_{0}}Z_{a_{1}}{}^{d_{0}}]\eta^{b_{0}}\eta^{a_{1}}\mathcal{P}_{a_{0}}\nonumber\\
&& + 2\frac{\partial \Omega^{(2)}}{\partial\mathcal{P}_{a_{0}}}G_{a_{0}} - 2\frac{\partial \Omega^{(2)}}{\partial\mathcal{P}_{a_1}}Z_{a_{1}}{}^{a_{0}}\mathcal{P}_{a_{0}}.
\end{eqnarray}
The use of the identities \eqref{identity-U}, \eqref{identity-D} and the reducibility relations \eqref{reducibility-chain1}, \eqref{reducibility-chain2} allows us to write
\begin{eqnarray}\label{omega2}
\Omega^{(2)} &=& -\frac{1}{3}U_{a_{0}b_{0}c_{0}}{}^{a_{1}}\eta^{a_{0}}\eta^{b_{0}}\eta^{c_{0}}\mathcal{P}_{a_{1}} - D_{b_{0}a_{1}}{}^{b_{1}}\eta^{b_{0}}\eta^{a_{1}}\mathcal{P}_{b_{1}} + Z_{a_{2}}{}^{a_{1}}\eta^{a_{2}}\mathcal{P}_{a_{1}} - \frac{1}{2}f_{a_{2}}{}^{a_{0}b_{0}}\eta^{a_{2}}\mathcal{P}_{a_{0}}\mathcal{P}_{b_{0}}. \nonumber\\
\end{eqnarray}
Finally one plugs eqns. \eqref{omega0}, \eqref{omega1}, \eqref{omega2} into \eqref{expression-for-B} to arrive at the formula
\begin{eqnarray}
B^{(2)} &=& M_{a_{0}b_{0}c_{0}d_{0}}{}^{a_{2}}Z_{a_{2}}{}^{a_{1}}\eta^{a_{0}}\eta^{b_{0}}\eta^{c_{0}}\eta^{d_{0}}\mathcal{P}_{a_{1}} + 2 S_{e_{0}b_{0}a_{1}}{}^{a_{2}}Z_{a_{2}}{}^{b_{1}}\eta^{e_{0}}\eta^{b_{0}}\eta^{a_{1}}\mathcal{P}_{b_{1}}\ - 2\eta^{b_{0}}\eta^{a_{2}}N_{e_{0}a_{2}}{}^{b_{2}}Z_{b_{2}}{}^{b_{1}}\mathcal{P}_{b_{1}}\nonumber\\
&& + \eta^{e_{0}}\eta^{a_{2}}\mathcal{P}_{a_{0}}\mathcal{P}_{b_{0}}\bigg(\partial_{e_{0}}f_{a_{2}}{}^{a_{0}b_{0}} - 2f_{a_{2}}{}^{f_{0}a_{0}}f_{f_{0}e_{0}}{}^{b_{0}}\bigg) - 2\eta^{a_{1}}\eta^{f_{1}}\mathcal{P}_{b_{1}}R_{a_{1}f_{1}}{}^{a_{2}}Z_{a_{2}}{}^{a_{1}}\nonumber\\
&& + \eta^{a_{1}}\eta^{b_{1}}\mathcal{P}_{a_{0}}\mathcal{P}_{f_{0}}[Z_{a_{1}}{}^{a_{0}}, Z_{b_{1}}{}^{f_{0}}] - \eta^{a_{1}}\mathcal{P}_{a_{0}}\eta^{f_{0}}\eta^{d_{0}}\mathcal{P}_{e_{0}}[Z_{a_{1}}{}^{a_{0}}, f_{f_{0}d_{0}}{}^{e_{0}}]\nonumber\\
&& + \frac{1}{4}\eta^{a_{0}}\eta^{b_{0}}\mathcal{P}_{c_{0}}\eta^{f_{0}}\eta^{d_{0}}\mathcal{P}_{e_{0}}[f_{a_{0}b_{0}}{}^{c_{0}}, f_{f_{0}d_{0}}{}^{e_{0}}] + 2\frac{\partial \Omega^{(3)}}{\partial \mathcal{P}_{a_{0}}}G_{a_{0}} - 2\frac{\partial \Omega^{(3)}}{\partial \mathcal{P}_{a_{1}}}Z_{a_{1}}{}^{a_{0}}\mathcal{P}_{a_{0}}\nonumber\\
&& + 2\frac{\partial\Omega^{(3)}}{\partial\mathcal{P}_{a_{2}}}\bigg(Z_{a_{2}}{}^{a_{1}}\mathcal{P}_{a_{1}} - \frac{1}{2}f_{a_{2}}{}^{a_{0}b_{0}}\mathcal{P}_{a_{0}}\mathcal{P}_{b_{0}}\bigg).
\end{eqnarray}
The identities \eqref{ma0b0c0d0a2}, \eqref{consistency-for-sa0b0a1a2}, \eqref{sa0b0a1a2}, \eqref{consistency-for-sa0b0a1a2}, \eqref{na0a2b2}, \eqref{consistency-for-na0a2b2}, \eqref{ra1b1a2}, \eqref{consistency-for-ra1b1a2} and the reducibility relations \eqref{reducibility-chain1}- \eqref{reducibility-chain3} then allows us to conclude that
\begin{eqnarray}
\Omega^{(3)} &=& -\frac{1}{2}M_{a_{0}b_{0}c_{0}d_{0}}{}^{a_{2}}\mathcal{P}_{a_{2}}\eta^{a_{0}}\eta^{b_{0}}\eta^{c_{0}}\eta^{d_{0}} - S_{e_{0}b_{0}a_{1}}{}^{a_{2}}\eta^{a_{1}}\eta^{e_{0}}\eta^{b_{0}}\mathcal{P}_{a_{2}}\nonumber\\
&& + N_{e_{0}a_{2}}{}^{b_{2}}\eta^{e_{0}}\eta^{a_{2}}\mathcal{P}_{a_{2}} + R_{d_{1}a_{1}}{}^{a_{2}}\eta^{d_{1}}\eta^{a_{1}}\mathcal{P}_{a_{2}}.
\end{eqnarray}
In this manner, the BRST operator for a 2-reducible gauge system reads
\begin{eqnarray}\label{generalBRSToperator}
Q &=& \eta^{a_{0}}G_{a_{0}} +  Z_{a_{1}}{}^{a_{0}}\eta^{a_{1}}\mathcal{P}_{a_{0}} - \frac{1}{2}f_{a_0 b_0}{}^{c_0}\eta^{a_0}\eta^{b_0}\mathcal{P}_{c_0} -\frac{1}{3}U_{a_{0}b_{0}c_{0}}{}^{a_{1}}\eta^{a_{0}}\eta^{b_{0}}\eta^{c_{0}}\mathcal{P}_{a_{1}}\nonumber\\
&& - D_{b_{0}a_{1}}{}^{b_{1}}\eta^{b_{0}}\eta^{a_{1}}\mathcal{P}_{b_{1}} + Z_{a_{2}}{}^{a_{1}}\eta^{a_{2}}\mathcal{P}_{a_{1}} - \frac{1}{2}f_{a_{2}}{}^{a_{0}b_{0}}\eta^{a_{2}}\mathcal{P}_{a_{0}}\mathcal{P}_{b_{0}}\nonumber\\
&& -\frac{1}{2}M_{a_{0}b_{0}c_{0}d_{0}}{}^{a_{2}}\mathcal{P}_{a_{2}}\eta^{a_{0}}\eta^{b_{0}}\eta^{c_{0}}\eta^{d_{0}} - S_{e_{0}b_{0}a_{1}}{}^{a_{2}}\eta^{a_{1}}\eta^{e_{0}}\eta^{b_{0}}\mathcal{P}_{a_{2}}\nonumber\\
&& + N_{e_{0}a_{2}}{}^{b_{2}}\eta^{e_{0}}\eta^{a_{2}}\mathcal{P}_{a_{2}} + R_{d_{1}a_{1}}{}^{a_{2}}\eta^{d_{1}}\eta^{a_{1}}\mathcal{P}_{a_{2}}.
\end{eqnarray}

\section{Relating the super-Lorentz generators}\label{AppendixB0}
In section \ref{section3} we claimed the super-Lorentz generators \eqref{sLorentzoriginal} and \eqref{definitionNps}, \eqref{definitionqtildeps} are related to each other via the maps \eqref{Lambda}, \eqref{Omega}, \eqref{psi}. In this Appendix we show explicitly how this does occur, and similar algebraic manipulations can be used to check all the others equations presented in this work. Let us start with the SUSY generator $q_{\alpha}$ given in eqn. \eqref{sLorentzoriginal}:
\begin{eqnarray}\label{auxeqn1}
q_{\alpha} &=& (\Lambda\gamma^{m})_{\alpha}\psi_{m}
\end{eqnarray}
After plugging eqns. \eqref{Lambda}, \eqref{psi} into \eqref{auxeqn1}, one finds
\begin{eqnarray}\label{qqtilde}
q_{\alpha} &=& \frac{2}{\gamma^{2}}(\lambda\gamma^{m})_{\alpha}\Gamma_{m} + \frac{1}{2(\lambda\nu)}(\lambda\gamma^{m})_{\alpha}(\nu\gamma_{m}\tilde{q}) + \frac{1}{2(\lambda\nu)}(\nu\gamma^{p}\gamma^{m})_{\alpha}(\lambda\gamma_{p}\bar{\lambda})\Gamma_{m}\nonumber\\
&& + \frac{\gamma^{2}}{8(\lambda\nu)^{2}}(\nu\gamma^{p}\gamma^{m})_{\alpha}(\nu\gamma_{m}\tilde{q})(\lambda\gamma_{p}\bar{\lambda})
\end{eqnarray}
where we have written down $\psi^{m}$ in the more convenient way
\begin{eqnarray}
\psi^{m} &=& \frac{2}{\gamma}\Gamma^{m} + \frac{\gamma}{2(\lambda\nu)}(\nu\gamma^{m}\tilde{q})
\end{eqnarray}
The first term in \eqref{qqtilde} vanishes because of eqn. \eqref{phialphaconstraint}, and the second term can be expressed as
\begin{eqnarray}\label{identity1}
\frac{1}{2(\lambda\nu)}(\lambda\gamma^{m})_{\alpha}(\nu\gamma_{m}\tilde{q}) &=& \frac{1}{2(\lambda\nu)}(\lambda\gamma_{m}\bar{\lambda})(\nu\gamma^{s}\gamma^{m})_{\alpha}\Gamma_{s} + \tilde{q}_{\alpha}
% + (\bar{\lambda}\Gamma^{m})_{\alpha}\Gamma_{m} - \frac{1}{2}(\lambda\gamma^{m})_{\alpha}\bar{\Gamma}_{m}
\end{eqnarray}
where eqn. \eqref{definitionqtildeps} was used. Thus, eqn. \eqref{qqtilde} becomes
\begin{eqnarray}
q_{\alpha} 
&=& \tilde{q}_{\alpha} + \frac{\gamma^{2}}{4(\lambda\nu)}\nu_{\alpha}(\nu\gamma^{m}\tilde{q})(\lambda\gamma_{m}\bar{\lambda})
\end{eqnarray}
Then, the use of eqns. \eqref{identity1}, \eqref{Bconstraint} allows us to conclude that
\begin{eqnarray}
q_{\alpha} 
&=& \tilde{q}_{\alpha} + \frac{\nu_{\alpha}\gamma^{2}}{2(\lambda\nu)}B
\end{eqnarray}

Finally, let us express $M^{mn}$ given in eqn. \eqref{sLorentzoriginal} in terms of the pure spinor twistor variables:
\begin{eqnarray}
M^{mn} &=& \frac{1}{2}(\Lambda\gamma^{mn}\Omega) + \psi^{m}\psi^{n}\nonumber\\
&=& \frac{1}{\gamma^{2}}\bigg[(\lambda\gamma^{mn}\mu) + 4\Gamma^{m}\Gamma^{n}\bigg] - \frac{1}{8(\lambda\nu)}\bigg[N^{pq}(\lambda\gamma^{mn}\gamma^{pq}\nu) + J_{\Omega}(\lambda\gamma^{mn}\nu)\bigg]\nonumber\\
&& + \frac{1}{4(\lambda\nu)}(\nu\gamma^{p}\gamma^{mn}\mu)(\lambda\gamma_{p}\bar{\lambda}) + \frac{1}{(\lambda\nu)}\Gamma^{m}(\nu\gamma^{n}\tilde{q}) + \frac{1}{(\lambda\nu)}(\nu\gamma^{m}\tilde{q})\Gamma^{n}\nonumber\\
&& + \frac{\gamma^{2}}{(\lambda\nu)^{2}}\bigg[-\frac{1}{32}(\nu\gamma^{spqmn}\nu)(\lambda\gamma_{s}\bar{\lambda})N_{pq} + \frac{1}{8}(\nu\gamma^{m}\tilde{q})(\nu\gamma^{n}\tilde{q})\bigg]
\end{eqnarray} 
The term proportional to $\gamma^{-2}$ vanishes on the support of eqn. \eqref{phimnconstraint}. The term inside the square brackets proportional to $\gamma^{0}$ can be cast as
%\begin{eqnarray}
%- \frac{1}{8(\lambda\nu)}\bigg[N^{pq}(\lambda\gamma^{mn}\gamma^{pq}\nu) + J_{\Omega}(\lambda\gamma^{mn}\nu)\bigg] &=& -\frac{1}{8(\lambda\nu)}\bigg[4N^{pq}(\lambda\gamma^{mq}\nu) - 4N^{mq}(\lambda\gamma^{nq}\nu) \nonumber\\
%&& + N^{pq}(\lambda\gamma^{pq}\gamma^{mn}\nu) + J_{\Omega}(\lambda\gamma^{mn}\nu)\bigg]
%\end{eqnarray}
\begin{eqnarray}
- \frac{1}{8(\lambda\nu)}\bigg[N^{pq}(\lambda\gamma^{mn}\gamma^{pq}\nu) + J_{\Omega}(\lambda\gamma^{mn}\nu)\bigg] &=& -\frac{1}{8(\lambda\nu)}\bigg[4(\lambda\gamma^{mn}\nu)K + 8(\tilde{q}\gamma^{m}\nu)\Gamma^{n} - 8(\tilde{q}\gamma^{n}\nu)\Gamma^{m}\nonumber\\
&& - 2(\lambda\gamma^{p}\bar{\lambda})(\nu\gamma^{m}\gamma_{p}\gamma^{n}\mu) + 2(\lambda\gamma^{p}\bar{\lambda})(\nu\gamma^{n}\gamma_{p}\gamma^{m}\mu)\nonumber\\
&& - 8(\lambda\nu)N^{mn} + J_{\Omega}(\lambda\gamma^{mn}\nu)\nonumber\\
&& -5(\lambda\bar{\mu})(\lambda\gamma^{mn}\nu) + 2(\lambda\gamma_{p}\bar{\lambda})(\mu\gamma^{p}\gamma^{mn}\nu)\nonumber\\
&& - (\bar{\lambda}\mu)(\lambda\gamma^{mn}\nu) - 2\bar{\Gamma}^{p}\Gamma_{p}(\lambda\gamma^{mn}\nu)\bigg]
\end{eqnarray}
where $K$ is given in eqn. \eqref{currentK}, and we have made use of the identity
\begin{eqnarray}\label{identity2}
(\lambda\gamma_{m})_{\alpha}N^{mn} &=& -\frac{1}{2}(\lambda\gamma^{n})_{\alpha}K + 2\tilde{q}_{\alpha}\Gamma^{n} - \frac{1}{2}(\lambda\gamma_{m}\bar{\lambda})(\gamma^{m}\gamma^{n}\mu)_{\alpha}
\end{eqnarray}
Therefore,
\begin{eqnarray}
M^{mn} &=& N^{mn} - \frac{1}{8(\lambda\nu)}\bigg[ 2(\lambda\gamma^{p}\bar{\lambda})(\mu\gamma_{p}\gamma^{mn}\nu) -4(\lambda\gamma^{m}\bar{\lambda})(\nu\gamma^{n}\mu) + 4(\lambda\gamma^{m}\bar{\lambda})(\nu\gamma^{n}\mu)\nonumber\\
&& + 2(\lambda\gamma^{p}\bar{\lambda})(\nu\gamma_{p}\gamma^{mn}\mu)\bigg] - \frac{\gamma^{2}}{32(\lambda\nu)^{2}}(\nu\gamma^{mnpqr}\nu)\tilde{M}_{pqr}
\end{eqnarray}
where $\tilde{M}^{mnp}$ is given in eqn. \eqref{PLconstraint}. Using standard gamma-matrix identities, the expression inside the square brackets can be shown to be identically zero, and so
\begin{eqnarray}
M^{mn} &=& N^{mn} - \frac{\gamma^{2}}{32(\lambda\nu)^{2}}(\nu\gamma^{mnpqr}\nu)\tilde{M}_{pqr}
\end{eqnarray}

\section{Review of Higher-Dimensional Twistor Transforms using Projective Pure Spinors}\label{AppendixB}
In this appendix we quickly review the main results of \cite{Berkovits:2004bw} which states that higher-dimensional twistor transforms are naturally realized by projective pure spinors. Let us start with the $D=2n$ massless Klein-Gordon equation of motion for a scalar field $\partial^{m}\partial_{m}\Phi(X)$ = 0, where $m = 1, \ldots, 2n$, which can be automatically solved by the identification
\begin{eqnarray}\label{map1}
\Phi(z,\bar{z}) &=& \oint\,du^{\frac{n(n-1)}{2}}\,f(u,v)|_{v_{j}=z_{j}+u_{jk}\bar{z}_{k}},
\end{eqnarray}
where $f(u,v)$ is a holomorphic function and $z$ is the complex coordinate defined through the relation $z_{j} = X_{2j-1} + iX_{2j}$, where $j = 1, \ldots, n$ and is related to $u,v$ via
\begin{eqnarray}\label{vu}
v_{j} &=& z_{j} + u_{jk}\bar{z}_{k},
\end{eqnarray}
with $u_{jk} = -u_{jk}$. The contour integral in \eqref{map1} is chosen arbitrarily.

In order to write eqn. \eqref{map1} in a Lorentz-covariant way, one needs to introduce projective pure spinors together with a well-defined measure on the space spanned by these ones. One then defines a projective pure spinor in $d=2n$ even dimension as a $2^{n-1}$-component chiral spinor $\lambda^{\alpha}$ satisfying the constraints
\begin{eqnarray}
\lambda\gamma^{m_{1}\ldots m_{n-4}}\lambda = 0 \hspace{2mm},\hspace{2mm}\lambda\gamma^{m_{1}\ldots m_{n-8}}\lambda = 0\hspace{2mm},\hspace{2mm}\lambda\gamma^{m_{1}\ldots m_{n-12}}\lambda = 0\hspace{2mm},\hspace{2mm}\ldots
\end{eqnarray}
where $(\gamma^{m})_{\hat{\alpha}\beta}$, $(\gamma^{m})^{\alpha\hat{\beta}}$ are the Pauli matrices of $SO(2n)$, and the identification
\begin{eqnarray}
\lambda^{\alpha} \sim C\lambda^{\alpha},
\end{eqnarray}
for some arbitrary complex parameter $C$. The antichiral spinor $w_{\hat{\alpha}}$ defined as
\begin{eqnarray}\label{wdef}
w_{\hat{\alpha}} &=& (\gamma^{m})_{\hat{\alpha}\beta}\lambda^{\beta}X_{m}
\end{eqnarray}
then satisfies the relations
\begin{eqnarray}
\lambda\gamma^{m_{1}\ldots m_{n-3}}w = 0 \hspace{2mm},\hspace{2mm} \lambda\gamma^{m_{1}\ldots m_{n-5}}w = 0 \hspace{2mm},\hspace{2mm} \lambda\gamma^{m_{1}\ldots m_{n-7}}w = 0\hspace{2mm},\hspace{2mm} \ldots
\end{eqnarray}
Note that $\hat{\alpha}$ and $\alpha$ in \eqref{wdef} belong to different irreducible spinor representations when $n$ is even, and belong to the same irreducible representation when $n$ is odd. Eqn. \eqref{wdef} is nothing but the Lorentz-covariant version of eqn. \eqref{vu}.

To write a well-defined measure on the projective pure spinor space, namely $SO(2n)/U(n)$, one introduces the tensor $T^{[\alpha_{1}\ldots \alpha_{R}](\beta_{1}\ldots \beta_{s})}$ which is fully antisymmetric in its first $R = 2^{n-1} - 1 - \frac{n(n-1)}{2}$ indices, and fully symmetric in its last $S = \frac{(n-2)(n-3)}{2}$ indices, and satisfies the constraints
\begin{equation}
(\gamma^{m_{1}\ldots m_{n-4}})_{\beta_{1}\beta_{2}}T^{[\alpha_{1}\ldots \alpha_{n}](\beta_{1}\beta_{2}\ldots \beta_{n})} = 0, \ \ \  (\gamma^{m_{1}\ldots m_{n-8}})_{\beta_{1}\beta_{2}}T^{[\alpha_{1}\ldots \alpha_{n}](\beta_{1}\beta_{2}\ldots \beta_{n})} = 0, \ \ldots
\end{equation}
An explicit way to construct this tensor is given in \cite{Berkovits:2004bw}. One can then use this object to define the Lorentz-invariant measure over the coset space $SO(2n)/U(n)$
\begin{eqnarray}\label{psmeasure}
[d\lambda]_{D=2n} &=& \frac{1}{(\lambda C)^{S}}\epsilon_{\alpha_{1}\ldots \alpha_{\frac{n(n-1)}{2}} \delta \beta_{1}\ldots \beta_{R}}d\lambda^{\alpha_{1}} \wedge\ldots\wedge d\lambda^{\alpha_{\frac{n(n-1)}{2}}}\lambda^{\delta}T^{[\beta_{1}\ldots \beta_{R}](\sigma_{1}\ldots \sigma_{S})}C_{\sigma_{1}}\ldots C_{\sigma_{S}},\nonumber\\
\end{eqnarray}
where $C_{\alpha}$ is a constant antichiral spinor. To see that \eqref{psmeasure} is Lorentz-invariant, one needs to show that \eqref{psmeasure} is independent of the choice of $C^{\alpha}$. As discussed in \cite{Berkovits:2004bw}, this immediately follows from the fact the only components of $C^{\alpha}$ which contribute in \eqref{psmeasure} are the ones which are $SU(n)$ singlets but $U(n)$ charged. Since the number of $C$'s in the numerator is the same as the ones appearing in the denominator, the measure \eqref{psmeasure} is independent of $C_{\alpha}$.

Therefore, the twistor transform formula \eqref{map1} written in Lorentz covariant form is given by
\begin{eqnarray}\label{psmap}
\Phi(X) &=& \int [d\lambda]_{D=2n} F(\lambda,w)|_{w_{\hat{\alpha}} = (\gamma^{m})_{\hat{\alpha}\beta}\lambda^{\beta}X_{m}},
\end{eqnarray}
where $F$ satisfies the homogeneity condition $F(h\lambda^{\alpha},hw_{\hat{\alpha}}) = h^{2-2n}F(\lambda^{\alpha},w_{\hat{\alpha}})$.

To describe massless $D=2n$ higher-spin fields, one easily generalizes \eqref{psmap} to the twistor transform formula
\begin{eqnarray}
\Phi(X)^{(\alpha_{1}\ldots \alpha_{N})} &=& \int [d\lambda]_{D=2n}\lambda^{\alpha_{1}}\ldots \lambda^{\alpha_{N}} F(\lambda,w)|_{w_{\hat{\alpha}} = (\gamma^{m})_{\hat{\alpha}\beta}\lambda^{\beta}X_{m}},
\end{eqnarray}
where $N$ is positive. Since $\lambda^{\alpha}$ is a pure spinor, it satisfies $(\gamma^{m})_{\hat{\alpha}\beta}(\gamma_{m})_{\hat{\delta}\epsilon}\lambda^{\beta}\lambda^{\epsilon} = 0$, and therefore
\begin{eqnarray}
(\gamma^{m})_{\hat{\alpha}\beta_{1}}\partial_{m}\Phi(X)^{\beta_{1}\beta_{2}\ldots \beta_{N}} = 0.
\end{eqnarray}


%\bibliographystyle{JHEP}
%\bibliography{references}

\providecommand{\href}[2]{#2}\begingroup\raggedright\endgroup

\end{document}